**Microbiome-derived bile acids contribute to elevated antigenic response and bone erosion in rheumatoid arthritis**

Xiuli Su[1,#], Xiaona Li[1,2,#], Yanqin Bian[3,#], Qing Ren[4], Leiguang Li[1], Xiaohao Wu[5], Hemi Luan[6], Bing He[5], Xiaojuan He[7], Hui Feng[3], Xingye Cheng[8], Pan-Jun Kim[1,9,10], Leihan Tang[1,8,10], Aiping Lu[5], Lianbo Xiao[3*], Liang Tian[1,8,10*], Zhu Yang[1*], Zongwei Cai[1*]

[1]State Key Laboratory of Environmental and Biological Analysis, Hong Kong Baptist University, Hong Kong, China

[2]Department of Pharmacy, Peking University Third Hospital, Beijing, China

[3]Institute of Arthritis Research in Integrative Medicine, Guanghua Hospital Affiliated to Shanghai University of Traditional Chinese Medicine, Shanghai, China

[4]Li Ka Shing Faculty of Medicine, University of Hong Kong, Hong Kong, China

[5]Law Sau Fai Institute for Advancing Translational Medicine in Bone and Joint Diseases, School of Chinese Medicine, Hong Kong Baptist University, Hong Kong, China

[6]SUSTech Academy for Advanced Interdisciplinary Studies, Southern University of Science and Technology, Shenzhen, China

[7]Institute of Basic Research in Clinical Medicine, China Academy of Chinese Medical Sciences, Beijing, China

[8]Department of Physics, Hong Kong Baptist University, Kowloon, Hong Kong, China

[9]Department of Biology, Hong Kong Baptist University, Kowloon, Hong Kong, China

[10]Institute of Computational and Theoretical Studies, Hong Kong Baptist University, Kowloon, Hong Kong, China

[#] These authors contributed equally.

*Corresponding author. E-mail: zwcai@hkbu.edu.hk (Z. Cai); zyang@hkbu.edu.hk (Z. Yang); liangtian@hkbu.edu.hk (L. Tian); xiao_lianbo@163.com (L. Xiao)

**Abstract**

Rheumatoid arthritis (RA) is a chronic, disabling and incurable autoimmune disease. It has been widely recognized that gut microbial dysbiosis is an important contributor to the pathogenesis of RA, although distinct alterations in microbiota have been associated with this disease. Yet, the metabolites that mediate the impacts of the gut microbiome on RA are less well understood. Here, with microbial profiling and non-targeted metabolomics, we revealed profound yet diverse perturbation of the gut microbiome and metabolome in RA patients in a discovery set. In the *Bacteroides*-dominated RA patients, differentiation of gut microbiome resulted in distinct bile acid profiles compared to healthy subjects. Predominated *Bacteroides* species expressing bile salt hydrolase (BSH) and 7α-hydroxysteroid dehydrogenase (HSDH) increased, leading to elevated secondary bile acid production in this subgroup of RA patients. Reduced serum fibroblast growth factor-19 and dysregulated bile acids were evidence of impaired farnesoid X receptor-mediated signaling in the patients. This gut microbiota-bile acid axis was correlated to anti-citrullinated peptide antibody (ACPA). The patients from the validation sets demonstrated that ACPA-positive patients have more abundant bacteria expressing BSH and 7alpha-HSDH but less *Clostridium scindens* expressing 7alpha-dehydroxylation enzymes, together with dysregulated microbial bile acid metabolism and more severe bone erosion than ACPA-negative ones. Mediation analyses revealed putative causal relationships between the gut microbiome, bile acids, and ACPA-positive RA, supporting a potential causal effect of *Bacteroides* species in increasing levels of ACPA and bone erosion mediated via disturbing bile acid metabolism. These results provide insights into the role of gut dysbiosis in RA in a manifestation-specific manner, as well as the functions of bile acids in this gut-joint axis, which may be a potential intervention target for precisely controlling RA conditions.

**Highlights**

- *Bacteroides* predominates in the gut microbiome of RA patients

- Microbial and circulating metabolic shifts correlate with the gut microbiome

- Perturbed bile acid metabolism mediates the association between the gut microbiome and antigenic response and bone erosion in RA

# 1 Introduction

Rheumatoid arthritis (RA) is a systemic autoimmune disease that could lead to deformity, unsteadiness and disability (Smolen et al., 2016). It often causes intensive pain, inflammation, and joint destruction. Both genetic and environmental factors play significant roles in the etiology of RA (Smolen et al., 2016). Accumulating evidence indicates that the gut microbiome serves as an important environmental factor related to the etiology of this disease (Lucas et al., 2018; Scher et al., 2013; Wells et al., 2020; Zhang et al., 2015). Alternations in the gut microbiota were observed in RA patients and partially normalized after therapy (Zhang et al., 2015). It was reported that microbial dysbiosis contributed to the development of joint inflammation in genetic arthritis-prone SKG mice (Maeda et al., 2016). Particularly, human leukocyte antigen alleles influence the predisposition to RA through interaction with the intestinal microbiome (Asquith et al., 2019; Wells et al., 2020). The microorganisms also regulate immune cell differentiation, inflammatory cytokine production, and intestinal permeability and barrier function, involved in the autoimmunity of RA (Taneja, 2017). *Bifidobacterium adolescentis* can alone induce T helper 17 cell accumulation in the small intestine and aggravate the autoimmune arthritis of K/BxN mice (Tan et al., 2016). Altogether, the gut microbiome has been shown to affect host immunity and the development of RA.

Emerging researches suggest that the gut microbiome and its products can affect immune homeostasis and host susceptibility to immune-mediated diseases and disorders (Rooks et al., 2016). Although the mechanisms have not yet been systematically studied, the alterations in the microbiota in RA likely result in changes in bacterial products across the intestinal barrier to act as crucial signals or priming stimuli for the innate immune system. The fecal metabolome provides a functional readout of microbial activities that mediate host-microbe interactions (Zierer et al., 2018). Besides, the human gut microbiota produces dozens of metabolites that can be transferred into the bloodstream, where they can have systemic effects on the host (Dodd et al., 2017; Wikoff et al., 2009). Thus fecal and serum metabolomics are useful tools to investigate the metabolic changes in response to endogenous or exogenous stimuli and the crosstalk between the intestinal microbiota and the host. For instance, immunomodulators such as short-chain fatty acids derived from microbial fermentation of dietary fibers in the intestine have been revealed to protect from inflammation-induced bone loss and inhibit osteoclast differentiation and bone resorption (Lucas et al., 2018). Nevertheless, there is still inadequate pathological proof in the respect of metabolic signatures for RA, particularly the metabolic perturbations correlated with the altered gut microbiome. Papers have rarely focused on fecal metabolomics in RA and illustrated the role of microbiome-derived metabolites.

Herein our study aimed to investigate RA-associated biomarkers and discover the association between metabolome perturbation and microbial dysbiosis under a rheumatic state. We combined microbial profiling and metabolomic profiling of fecal and serum samples from RA subjects and healthy control (HC) individuals in a 94-member discovery set (Fig. 1). One of our major observations was that RA-elevated *Bacteroides* enriched secondary bile acid biosynthesis and showed positive links with the antigenic response in RA. In the other two independent validation sets, targeted quantification of bile acids and shotgun metagenomic sequencing demonstrated the associations among the gut microbiome, secondary bile acids, antigenic response and bone erosion in RA, validating the previous findings. Our study identified the role of bile acid metabolism in host-microbiome interactions of RA patients, especially in the ones whose gut microbiotas were dominated by *Bacteroides* genus, making our findings a useful resource for the study of potential intervention targets for RA.

## 2 Results

### 2.1 Gut microbiome dysbiosis in response to RA

We investigated the microbial compositions of fecal samples from 44 RA patients and 50 healthy subjects by using 16S rRNA amplicon sequencing (Supplementary Table 1, Supplementary Fig. 1A-B). Canonical correspondence analysis (Supplementary Fig. 1C) showed that the gut microbiome of RA and HC groups separated along the vector for RA status, confirming that microbial variations were associated mainly with RA status rather than with other covariates (age, sex, BMI). The principal coordinates analysis (PCoA) at the genus level was applied to examine the principal driving factors of microbial variance (Fig. 2A). The plot of PCoA revealed a significant dissimilarity of microbial composition between the healthy control (HC) and RA groups (PERANOVA, $p < 0.001$), mainly explained by PCo-2 dimension. PCo-2 dimension was significantly associated with the abundances of the phyla Firmicutes and Bacteroidetes, which explained about 23.4% of sample variations. We observed that RA individuals had more Bacteroidetes ($p < 0.001$) and fewer Firmicutes ($p < 0.01$) than control individuals (Supplementary Fig. 1A, Supplementary Fig. 1D), which was consistent with the previous reports (Jeong et al., 2019; Li et al., 2021). The first principal coordinate (PCo-1, explaining 35.9% variation), however, was strongly correlated with the abundances of two Bacteroidetes genera, *Prevotella* and *Bacteroides* (Fig. 2A), which have been defined as two major indicators of human gut enterotypes (Costea et al., 2018). *Bacteroides* and *Prevotella*, predominated in the gut microbiome of two subgroups of RA subjects, respectively, contributing more variation among samples than the difference between the RA and control individuals. As reported in the previous studies (Manor et al., 2020; Wu et al., 2011), *Bacteroides*-dominated and *Prevotella*-dominated constitutions were found to be relatively non-overlapping (Fig. 2A). Previous studies have identified both *Prevotella*-rich and *Bacteroides*-rich microbial composition in fecal samples from the RA patients (Maeda et al., 2016; Scher et al., 2013). In addition, linear discriminant analysis (LDA) confirmed that *Bacteroides* and *Prevotella* were the top two genera that significantly contributed to the distinction between RA and HC (Fig. 2B). The overgrowth of *Prevotella* has been identified as a signature of microbial dysbiosis in new-onset untreated RA patients, which results in enhanced susceptibility to arthritis, increased osteoclast numbers and inflammation-induced bone loss (Kishikawa et al., 2020; Lucas et al., 2018; Maeda et al., 2016; Scher et al., 2013). A higher prevalence of *Bacteroides* was also detected in RA patients (Rodrigues et al., 2019; Sun et al., 2019). In addition to *Bacteroides* and *Prevotella*, *Parabacteroides, Phascolarctobacterium*, *Lactobacillus,* and *Paraprevotella* enriched in RA patients, whereas *Faecalibacterium*, *Escherichia-Shigella*, *Pseudobutyrivibrio*, *Roseburia, Anaerostipes*, and *Enterobacter* were found to be less abundant in RA (Fig. 2B). Among these, *Pseudobutyrivibrio* was previously reported to be negatively associated with psoriatic arthritis (Scher et al., 2015). The control-enriched genera *Roseburia* and *Faecalibacterium* were producers of butyrate, a metabolite involved in the regulation of immune and bone homeostasis (Lucas et al., 2018; Machiels et al., 2014; Rosser et al., 2020).

Differences in the taxonomic composition of the two groups led to varying community properties, such as richness and evenness. RA individuals presented lower alpha diversities (Shannon index, ACE index, Chao index, Fisher index and Simpson index, $p < 0.0001$, Fig. 2C, Supplementary Fig. 1E) than control individuals, suggesting that bacterial richness and evenness were lower in RA patients than those in control individuals. The reduction in microbial richness was found in both *Bacteroides*- and *Prevotella*-dominated RA patients (Fig. 2D). In contrast, bacterial *Bacteroides*/*Prevotella* ratio changed in the subjects, independent of the disease conditions (Fig. 2E). In the RA patients seropositive in anti-citrullinated peptide antibody (ACPA), we found that *Bacteroides* positively correlated with serum ACPA (Fig. 2F). Together, these findings revealed that the gut microbiome of RA individuals showed great dissimilarity to healthy individuals but the gut microbial changes in RA were strongly correlated with the RA manifestations, suggesting potential roles of microbial shifts in this disease.

**2.2 Fecal and serum metabolic shifts related to the gut microbiome**

Because gut lumen metabolites and circulating metabolomes represent a distinctive avenue for probing microbiome-host interaction, we performed nontargeted metabolomics on both fecal and serum samples from 94 subjects in the discovery set to examine alterations of microbiota-associated metabolites in response to RA. Metabolic features were detected based on liquid chromatography-mass spectrometry (LC-MS) and gas chromatography-mass spectrometry (GC-MS) platforms, then annotated by using an in-house library and public spectral databases (see details in Methods). A total of 467 and 292 compounds were identified with a metabolomics standards initiative level 1 or 2 identification (Viant et al., 2017) in fecal and serum metabolic profiling, respectively (Supplementary Table 2-3). Of them, 164 compounds overlapped in the two sample types (Supplementary Fig. 2A). The metabolic profiling of feces and serum distinguished RA individuals from healthy controls, as illustrated in the partial least squares discriminant analysis (PLS-DA) score plot (Fig. 3A-B), suggesting metabolic changes occurred during RA. To investigate the extent to which the gut microbiome was related to fecal and circulating metabolites in the host, we evaluated the inherent correlations between microbiome and metabolome in the two-way orthogonal partial least squares (O2PLS) model. Both metabolomes in fecal and serum samples showed significant correlations with the gut microbiome (Supplementary Fig. 2B-C), thereby supporting the notion that microbial and circulating metabolomes provide a functional readout of microbial activities and disease.

Next, we examined the overall pathway changes of fecal and serum metabolites through MetaboAnalyst (Chong et al., 2018). The results of altered pathways revealed that metabolic shifts were distinct between fecal and serum samples (Fig. 3C). The top altered microbial pathways in feces and serum included bile acid biosynthesis, tryptophan metabolism and aminoacyl-tRNA biosynthesis. Interestingly, accumulating evidence has implicated that bile acids and tryptophan catabolites produced by the gut microbiome are crucial mediators in the cross-talk between the host and the gut microbiome (Jia et al., 2018; Roager et al., 2018). To further investigate the role of microbial bile acid metabolism in the disease, we performed targeted quantification using LC-MS/MS to get a thorough profile of bile acids (Fig. 3D-F). No difference in the total fecal and serum bile acids was observed between RA and HC groups (Fig. 3D). Nevertheless, the constitution of the fecal bile acid pool differed between the groups. Patients with RA had lower proportions of conjugated primary bile acids and conjugated secondary bile acids but a higher percentage of unconjugated secondary bile acids in feces (Fig. 3E, Fig. 3G). Reduced level of serum fibroblast growth factor-19 (FGF-19) (Fig. 3H), a regulator of bile acid metabolism, suggested evidence of impaired farnesoid X receptor (FXR)-mediated signaling in RA (Jiao et al., 2018).

The observed metabolic alterations in RA patients evidenced that different microbial communities have varying metabolic properties and the microbiome is one of the strongest determinants of circulating metabolomes (Bar et al., 2020). The dominant *Bacteroides* is well known to involve in bile acid biosynthsis (Jia et al., 2018). Bile acids were highlighted in both fecal and serum metabolome (Fig. 3C, Supplementary Table 2-3), inspiring our subsequent study on bile acids. The bacterial genera expressing bile salt hydrolase (BSH), which is involved in the deconjugation of glyco-conjugated or tauro-conjugated bile acids, included *Bacteroides*, *Clostridium*, *Bifidobacterium* and *Lactobacillus* (Jia et al., 2018). After that, the unconjugated bile acids were modified via 7α-dehydroxylation catalyzed by multiple enzymes to produce deoxycholic acid (DCA) and lithocholic acid (LCA), or by hydroxysteroid dehydrogenase (HSDH) to produce iso-, allo- and oxo-bile acids. We found an increase in the abundance of the bacterial genera expressing BSH and 7α-HSDH (Fig. 3I), leading to the reduction of fecal conjugated bile acids, such as glycocholic acid (GCA), glycochenodeoxycholic acid (GCDCA), glycohyocholic acid (GHCA), taurodeoxycholic acid (TDCA) and taurolithocholic acid (TLCA) (Fig. 3J). Decreased circulating LCA (Figure 3K) suggested downregulated 7α-dehydroxylation in the microbiome of RA patients, although *Clostridium scindens*, which is known to possess 7α-dehydroxylation capacity (Sato et al., 2021), was not detected in our 16S rRNA amplicon sequencing due to its low resolution. In brief, these results indicated that the altered gut microbiota, especially the augment of *Bacteroides* in RA individuals, was associated with bile acids in feces and host circulation.

In addition, the gut microbiome in RA patients was prone to produce more indolelactate, 1H-indole-3-carboxaldehyde and 3-indoleacetic acid rather than 3-indolepropionic acid (Supplementary Fig. 2D). The augment of bacterial producers of indolelactate, 1H-indole-3-carboxaldehyde and 3-indoleacetic acid (Roager et al., 2018) in RA group increased the level of fecal 1H-indole-3-carboxaldehyde and 3-indoleacetic acid as well as circulating indolelactate (Supplementary Fig. 2E). On the other hand, the producers of 3-indolepropionic acid and were decreased in RA group. In sum, the present results highlighted that microbial modulation of metabolome was pivotal to regulating RA-induced metabolic disorders.

**2.3 Gut microbiome and microbiome-derived bile acids associated with clinical indices**

To ascertain the associations between clinical indices and microbial alterations, we performed boosted additive generalized linear models. After adjusting age, sex and BMI, we found that *Bacteroides* positively correlated with ACPA, but negatively associated with rheumatoid factors (RF) (IgM and IgG) (Fig. 4A). Additionally, *Bacteroides* was negatively associated with the levels of erythrocyte sedimentation rate (ESR), a marker of inflammation in the body. On the contrary, RA-enriched *Prevotella* was positively associated with RF IgM and RF IgG but not with ACPA. Furthermore, *Prevotella* is positively associated with ESR. These results suggested that *Bacteroides*- and *Prevotella*-rich gut microbial composition may be associated with distinct antigenic responses of ACPA and RF in the host. Intriguingly, we also found that genera expressing BSH and HSDH showed positive correlations with ACPA (Fig. 4B). In addition, we observed that ACPA was associated with increased fecal hyodeoxycholic acid (HDCA), serum taurohyodeoxycholic acid (THDCA) and glycoursodeoxycholic acid (GUDCA) (Fig. 4C). Conversely, ACPA negatively associated with 12-ketolithocholic acid (12-oxoLCA), 3-dehydrocholic acid (3-oxoCA), dehydrochenodeoxycholic acid (3-oxoCDCA), isodeoxycholic acid (isoDCA), isolithocholic acid (isoLCA), DCA, LCA, 23-nordeoxycholic acid (NorDCA) and glycolithocholic acid (GLCA), indicating that bile acids may be potential biomarkers linking gut microbiota and ACPA in RA patients. Moreover, norcholic acid (NorCA), which is an FXR antagonist (Gong et al., 2021), positively correlated with clinical indicators, including RA duration, C-reactive protein (CRP), disease activity score 28 (DAS28), ESR, RF IgM, RF IgG and RF IgA. Fecal GCA is negatively associated with DAS28(3). These findings highlight that the microbiome and its derived bile acids underlay the clinical characteristics of RA patients.

**2.4 Microbial bile acids contributed to elevated ACPA expression and bone erosion in RA**

To validate the association between the gut microbiome and ACPA in RA, we performed shotgun metagenomic sequencing and targeted bile acids profiling of fecal and/or serum samples from the validation set 1, as well as targeted bile acids analysis of serum samples from the validation set 2 (Fig. 1, Supplementary Table 1). We found that ACPA+ RA individuals had lower bacterial diversities than both undifferentiated arthritis (UA) and ACPA- RA groups (Fig. 5A, Supplementary Fig. 3), indicating a reduced bacterial richness in ACPA+ RA group. We also found higher β diversity of the gut microbiome in ACPA+ RA patients, suggesting a more heterogeneous microbial community structure among ACPA+ RA individuals than those in ACPA- RA and UA groups (Fig. 5B). Taxonomy analysis revealed that ACPA+ RA had a higher prevalence of *Bacteroides* genus compared to UA and ACPA- RA group as expected (Fig. 5C). Furthermore, species of *Bacteroides*, including *B. fragilis*, *B. ovatus*, *B. thetaiotaomicron* and *B. dorei*, were enriched in ACPA+ RA (Fig. 5D-E). Notably, these species harbored genes encoding BSH enzymes, which led to the reduction of GCDCA and GLCA (Fig. 5F, Supplementary Fig. 4-5). In addition, *B. fragilis* and *B. thetaiotaomicron* harbored genes encoding 7α-HSDH enzymes which produce 7-ketolithocholic acid (7-oxoLCA) from chenodeoxycholic acid (CDCA) (Bennett et al., 2003). Along with the augment of *Bacteroides* species, we observed a significant decrease in *Clostridium scindens* (Fig. 5E), indicating downregulated 7α-dehydroxylation in ACPA+ RA group.

It was noteworthy that ACPA+ RA patients had higher levels of CDCA, 7-oxoLCA, ursodeoxycholic acid (UDCA), tauroursodeoxycholic acid (TUDCA) and isoursodeoxycholic acid (isoUDCA) but lower amounts of dehydrolithocholic acid (3-oxoLCA), 3-oxoDCA, 3-oxoCDCA, LCA, DCA, isoDCA, isoLCA and 12-oxoLCA (Fig. 5F, Supplementary Fig. 4-5). It seemed that the gut microbiome in ACPA+ RA patients was prone to modify CDCA through 7α/β-epimerization rather than 7α-dehydroxylation or 3α- dehydrogenation. The downregulated 7α-dehydroxylation of CA, CDCA and UDCA resulted in the abatement of downstream bile acids. On the other hand, NorCA concentration increased in ACPA+ RA. These observations suggested that low-richness gut microbiome in ACPA+ RA patients showed the distinct metabolic potential of bile acids compared to UA and ACPA- RA groups. However, there is no significant difference in serum FGF-19 between ACPA+ and ACPA- groups (Fig. 5G).

We quantified the concentration of peptidylarginine deiminase (PAD) enzymes that generate the citrullinated protein targets of ACPA (Curran et al., 2020). Serum levels of PAD4 and PAD2 were higher in ACPA+ RA group than ACPA- RA group in the validation set 2 as expected (Fig. 5H-I). To investigate the impact of seropositive ACPA, we reviewed radiographic data of RA patients of ACPA- and ACPA+ RA patients. ACPA+ RA patients had more severe bone erosion and narrowing/(sub)luxation estimated in the light of the modified Sharp/van der Heijde score (SHS) score method (Liang et al., 2019) (Fig. 5J-K). Additionally, the scores of bone erosion and narrowing/(sub)luxation positively correlated with PAD4 but not PAD2 (Supplementary Fig. 6).

To systematically study putative causal relationships between *Bacteroides*, bile acids and disease phenotypes, we performed mediation analysis based on the validation sets. This approach established mediation links between the impact of *Bacteroides* species on ACPA, PAD2, PAD4 and SHS through bile acids ($p < 0.05$; Fig. 6A, Supplementary Table 4). For instance, *B. fragilis* can affect the expression of ACPA and PAD2 mediated by 12-oxoLCA, 3-DCA, GLCA and isoUDCA (Fig. 6B). We also observed that *B. thetaiotaomicron* negatively impacts isoLCA, which in turn increases ACPA expression (Fig. 6C). Additionally, *C. scindens* affected SHS through the mediation of LCA (Fig. 6D). In sum, these results supported a potential causal relationship between the gut microbiome and host phenotypes that was mediated by bile acids.

## 3 Discussion

Growing interest has been shown in the role of microbiome-driven metabolic reorganization and immunologic dysfunction in RA progression (Lucas et al., 2018; Rosser et al., 2020; Takahashi et al., 2020). We found that RA status was accompanied by microbial dysbiosis, including high prevalences of *Bacteroides* and *Prevotella*, and abatement of Firmicutes genera (*i.e.*, *Roseburia* and *Faecalibacterium*). *Bacteroides* and *Prevotella* are major microbes in the human gut microbiome (Costea et al., 2018). Previous research has found that direct transformations between *Bacteroides*-rich and *Prevotella*-rich communities within individuals were rare (Levy et al., 2020), indicating interactions between these taxa with the host phenotypes form their specific linkages. Low-richness of *Bacteroides*-dominated microflora had reduced functional redundancy, potentially suggesting a decreased resilience to perturbation, in accord with its frequent correlation with dysbiosis (Vieira-Silva et al., 2016). The ecology of *Bacteroides* affects its host by modulating immunity and metabolism in many diseases, such as obesity (Liu et al., 2017), orchestrates polycystic ovary syndrome (Qi et al., 2019), lethal inflammatory cardiomyopathy (Gil-Cruz et al., 2019) and familial adenomatous polyposis (Dejea et al., 2018). For instance, *B. thetaiotaomicron* encodes a cross-reactive β-galactosidase mimetic peptide that activates myosin-specific T cells in the gut, triggering autoimmunity with cardiomyopathy (Gil-Cruz et al., 2019). Augment of *Bacteroides* species in RA patients was also found in a previous study of another Chinese cohort (Zhang et al., 2015).

In this study, we performed untargeted metabolomics profiling and identified metabolic changes in the feces and serum of RA patients, including bile acids and tryptophan catabolites. We detected many of the RA-associated metabolites revealed in previous studies (such as amino acids, lipids, and nucleotides) (Guma et al., 2016; Tong et al., 2020), validating our approach. At the same time, we also noted that secondary bile acids identified in our study were not well-studied previously. Here, we depicted the detailed profile of bile acids in feces and serum in RA patients by targeted metabolomics. In addition, we discovered secondary bile acids in RA, such as UDCA, isoDCA and isoLCA, that potentially affect host autoimmunity in RA patients. These secondary bile acids modulate peripheral regulatory T cells (Hang et al., 2019; Shen et al., 2022; Song et al., 2020) which may contribute to the overactivity of host immune cells. For instance, UDCA significantly restrains the differentiation and activation of Treg cells and finally abates Treg-mediated immunosuppression (Shen et al., 2022). In the meantime, isoDCA potentiates the differentiation of peripherally induced Treg cells (Song et al., 2020). Moreover, 3-oxoLCA and isoLCA could inhibit the differentiation of Th17 cell (Hang et al., 2019). The accumulation of UDCA along with the reduction of isoDCA, isoLCA and 3-oxoLCA indicated dysbiosis of peripheral regulatory T cell modulators in RA patients. Intriguingly, these metabolite changes were presumably in line with important microbial biological properties during RA development. The abatement of conjugated primary bile acids, and 3/12-oxo- or epi-bile acids as well as augment of UDCA were in coordination with the accumulation of *Bacteroides* species and decrease of *C. scindens*. In addition, NorCA is an antagonist of FXR which is a bile acid receptor essential for maintaining the intestinal barrier and immunity (Gadaleta et al., 2011; Gong et al., 2021; Vavassori et al., 2009). However, the knowledge of its biosynthetic genes and microbial producers is not well understood. On the other hand, tryptophan-derived metabolites are ligands implicated in the activation of aryl hydrocarbon receptor to induce increased production of interleukin-22, which enhances inflammatory responses in RA synovial tissues as well as induces the proliferation and chemokine production of synovial fibroblasts (Ikeuchi et al., 2005). By contrast, the circulating 3-indolepropionic acid, which had anti-inflammatory and antioxidative effects (Roager et al., 2018), was downregulated. Notably, the abatement of tryptophan and the accumulation of kynurenine in the host reflected the activation of indoleamine-2,3-dioxygenase (IDO) enzyme. IDO-mediated degradation of tryptophan has been regarded as an important feedback mechanism modulating overactive immune responses, a signature of autoimmune diseases (Platten et al., 2019).

These changes in the gut microbiome and microbiome-derived metabolites, especially copious *Bacteroides* and deregulated bile acids, presumably accommodate or reflect host antigenic response and bone damage during RA progress. Our study revealed the augment of *Bacteroides* was positively correlated with ACPA and negatively associated with RF-IgM. Furthermore, ACPA+ RA patients have more abundant *Bacteroides* species expressing BSH and 7α-HSDH alongside a lower level of *C. scindens* expressing 7α-dehydroxylase. This caused the accumulation of UDCA derivatives (7-oxoLCA, TUDCA, isoUDCA) and the reduction of 3/12-oxo- or epi-bile acids (*e.g.* isoLCA, 12-oxoLCA, 3-oxoCDCA). RA is a heterogeneous disease and has two principal subsets characterized by the seropositive of ACPA and RF (Malmström et al., 2017). ACPA and RF often exist in the blood preceding any symptoms of joint inflammation. Their etiology in RA is different regarding both genetic and environmental factors (Malmström et al., 2017). Few patients seroconvert after RA occurrence, indicating that the observed antibody responses are more likely a trigger than an outcome of RA. Nevertheless, ACPA is more specific for RA since RF also exists in patients with other diseases (Malmström et al., 2017). Furthermore, mediation analysis indicated that these microbiome-derived bile acids mediated the impact of *Bacteroides* species on host ACPA expression and the severity of bone damage. Previous researches have revealed that patients with osteoporosis had a higher prevalence of *Bacteroides* than healthy subjects (Ling et al., 2021; Wei et al., 2021). At the same time, a previous study has found that bile acids promote neutrophil extracellular traps (NETs) formation via PAD4 and ROS (Zhang et al., 2016). And NETs might be a source of citrullinated autoantigens in RA (Wright et al., 2014). Additionally, activation of bile acid receptors, FXR and G protein-coupled bile acid receptor (TGR5), could prevent estrogen-dependent bone loss and enhance osteoblastogenesis in mice (Li et al., 2019). Besides, previous findings revealed that ACPA binds specifically to osteoclasts and their precursors in the normal bone and joint compartment (Catrina et al., 2017). Osteoclast differentiation depends on the citrullination of proteins by PAD (Catrina et al., 2017).

In summary, the combination of 16S amplicon, shotgun metagenomics, untargeted and targeted metabolomics revealed the landscape of gut microbiome and microbial metabolome changes during RA development with high resolution. To the best of our knowledge, this is the first study to report the role of *Bacteroides* and microbiome-derived bile acids in RA. We provided a landscape of less well-studied secondary bile acids in RA. Despite the relatively small sample size, we also demonstrated that microbiome-derived bile acids may contribute to host immunity and bone phenotype in subjects with RA in validation cohorts. The finding can provide insight for future research on RA treatment.

**4 Material and methods**

**4.1 Clinical sample collection**

In the study, we collected serum and/or stool samples from subjects in three independent cohorts including 174 RA patients, 6 UA patients and 110 healthy subjects recruited from Guanghua Integrative Medicine Hospital, Shanghai, China (Fig. 1, Supplementary Table 1). Healthy volunteers were enrolled by routine physical examination. We recruited individuals with RA aged between 18 and 75 years old who were diagnosed as RA with a disease duration of at least 6 months and met the American College of Rheumatology (ACR) criteria and the selection criteria. Seronegative UA patients with recent onset arthritis of at least one joint and symptom for 6-12 months were consecutively included. Given that the gut microbiome and metabolites vary with many different factors, including diet, diseases and medication, subjects were selected according to the criteria as follows to balance the baseline of the two groups. Individuals with different habits (*e.g.* complete vegetarian diet, alcohol or yogurt consumption) that may influence gut microbiome were excluded. In addition, individuals who suffered from other diseases, such as kidney, liver, lung, cardiovascular, hematologic disease, other autoimmune diseases, mental diseases and metabolic diseases, were excluded, especially individuals who were experiencing gastrointestinal diseases. Women who were pregnant, breastfeeding, or planning to be pregnant were excluded. Furthermore, individuals continuously receiving medications, especially disease-modifying antirheumatic drugs, nonsteroidal anti-inflammatory drugs, corticosteroids or painkillers for over six months, and individuals who received the aforementioned medicine within one month were excluded. What is more, individuals who ever received treatment of probiotics or antibiotics within two months were excluded. Clinical information of subjects was collected upon their first visit to the hospital according to standard procedures. Serum and fecal samples were quenched within 30 min after collection. All samples were stored at -80 °C until analysis. Informed consent was obtained from all participants or their legal representatives before sample collection. Questionnaires including diet information, dietary habits, smoker/nonsmoker status and basic treatment were conducted. The study was approved by the Ethics Committee of Institute of Basic Research in Clinical Medicine, Beijing, China.

**4.2 Human fecal microbiota analysis**

For 16S rRNA amplicon sequencing, total DNA was extracted using an E.Z.N.A.Soil DNA Kit (Omega, USA) according to the manufacturer's instructions. KAPA HiFi Hot Start Ready Mix (2×) (TaKaRa Bio Inc., Japan) was used to amplify the 16S rRNA V3-V4 amplicon. Sequencing was performed based on Illumina MiSeq system (Illumina MiSeq, USA). All effective bacterial sequences were submitted for subsequent analysis. Raw data were processed according to a previously reported method (Wilck et al., 2017).

For shotgun metagenomics sequencing, DNA exaction, single-end metagenomics sequencing and data processing were performed as previously described (Nie et al., 2022). The relative abundances of genera or species were used in the current study.

### 4.3 Metabolic profiling of human samples

Fecal metabolites were extracted via sonication from freeze-dried feces with chilled water and methanol. Serum metabolites were extracted from 50 μL serum through deproteinizing with 200 μL methanol containing internal standards. After centrifugation, the supernatants were collected and lyophilized. The residue was redissolved for the subsequent global metabolomics analysis using ultra-high liquid chromatograph-mass spectrometer and gas chromatography-mass spectrometry according to a previously reported method (Lu et al., 2019). Quality control (QC) samples were prepared by pooling an equal volume of each sample together to monitor the repeatability and stability of LC-MS analysis. The samples were analyzed randomly. MS data were processed according to a previously reported method (Luan et al., 2015). The metabolic features were annotated by comparing accurate mass, retention time, MS/MS pattern and isotope pattern of metabolites in QC samples with those of authentic standards or databases, *e.g.*, METLIN (www.metlin.scripps.edu), HMDB (www.hmdb.ca/), mzCloud

(www.mzcloud.org), MassBank of North America (mona.fiehnlab.ucdavis.edu/), *etc*. Enrichment analysis was performed through MetaboAnalyst 3.0 (www.metaboanalyst.ca/). In consideration of human and microbial metabolomes, enrichment analysis was done using a pathway library built based on that of homo sapiens and bacteria from KEGG database.

**4.4 Targeted analysis of bile acids**

Methanol containing internal standards (cholic acid-d4, deoxycholic acid-d4, glycocholic acid-d4, lithocholic acid-d4) was added to serum, then vortex was applied to extract the metabolites. Freeze-dried feces were homogenized in a mixture of chilled methanol/water (4:1, v/v) and internal standards. After centrifugation, the supernatants were lyophilized. The residues were dissolved for subsequence analysis.

Fifty-four bile acids (Supplementary Table 5) in serum and fecal samples were quantified by Ultimate 3000 RSLC coupled to QuantivaTM triple-quadrupole MS (Thermo Scientific, USA). An ACQUITY UPLC BEH C18 column (100 mm × 2.1 mm, 1.7 μm, Waters, USA) was applied for metabolite separation at 45 °C. The mobile phases were water and acetonitrile both containing 0.01% formic acid. The flow rate was 0.35 mL/min. The LC gradient program was as follows: 0-0.5 min, 25 % B; 14 min, 40 % B; 25 min, 67.5 % B; 25.5-28.5 min, 100 % B. The MS was equipped with a heated electrospray ionization source in negative ion mode. The spray voltage was 2.6 kV. The pressure of the sheath and auxiliary gas was set at 30 arb and 10 arb. The ion transfer tube temperature and vaporizer temperature were 300 °C. The CID gas was set at 2 mTorr.

**4.5 Quantification of serum antibodies**

Serum PAD2 and PAD4 concentrations were measured using commercial ELISA kits from Cayman Chemical according to the manufacturer's instructions. Serum FGF-19 concentration was measured with commercial ELISA kits from Thermo Fisher Scientific.

**4.6 Radiological evaluation by modified Sharp's method**

The radiological evaluation of the hands was done by two qualified doctors. The readers of the radiographs were blind to the order of the radiographs and the clinical information of RA patients. The scores of bone erosion and narrowing/(sub)luxation were estimated according to the modified Sharp/van der Heijde score (SHS) score method (Liang et al., 2019).

**4.7 Statistical analysis**

Statistical analysis was performed using R (v3.6) and SPSS. Data distribution was tested by using Shapiro-Wilk test. The significance of differential gut microbiome and metabolome was examed by a two-tailed Wilcoxon rank-sum test between HC and RA groups. Differential abundance of the gut microbiome and metabolites in ACPA+ RA, ACPA- RA and UA/HC individuals was assessed by one-way ANOVA. Adjusted $p$ values were calculated by Benjamini-Hochberg correction for multiple tests. Associations between the clinical index and the gut microbiome as well as metabolomes were assessed by boosted additive generalized linear models using MaAsLin2 package. Mediation analysis was done based on mediation R package.

**Acknowledgements**

This work was financially supported by Interdisciplinary Research Matching Scheme (IRMS) of Hong Kong Baptist University (Project ID RC-IRMS/13-14/03). L.X. greatly thanks to Hong Kong PhD Fellowship Scheme (HKPFS).

**Author contributions**

S.X., L.X., and B.Y. contributed equally to this work. S.X. wrote the manuscript and performed the majority of the experiments and data analyses. L.X. conducted the metabolic profiling. B.Y., R.Q., W.X., H.B. and F.H. helped collect samples and read the radiographs. L.L. contributed to the metabolic experiments. L.H. and C.X. assisted with the data analysis. H.X., K.P. and T.LH. offered suggestions on the study. C.Z., L.A., X.L., Y.Z. and T.L. initiated and supervised the study.

**Competing interests**

The authors declare that they have no conflict of interest.

**Figure legend**

**Figure 1** Overview shows the study setting. ACPA, anti-citrullinated peptide antibody; BA, bile acid; HC, healthy control; RA, rheumatoid arthritis; UA, undifferentiated arthritis.

**Figure 2** Gut microbial alterations between rheumatoid arthritis (RA) and healthy control (HC) groups. (A) Principal Coordinates Analysis (PCoA) at genus level based on the distance method of Jensen-Shannon Divergence. Each point represents one sample. Scatter plots of the relative abundances of the phyla Firmicutes and Bacteroidetes (PCo2) and the genera *Bacteroides* and *Prevotella* (PCo1) in each sample are shown across the corresponding principal component. (B) Linear discriminant analysis (LDA) scores of differentially abundant genera by LDA Effect Size. (C) Shannon index of the two groups estimating the number of OTUs in each sample gradually changes over the ordination of samples on Jensen-Shannon distance transformed space. (D) Shannon diversity index. Median values, the first and third quartiles are represented by solid lines (Mann-Whitney test, ***$p < 0.001$). (E) *Bacteroides*/*Prevotella* ratio (B/P ratio) in each sample over the ordination of samples on Jensen-Shannon distance transformed space. (F) The correlation between *Bacteroides* abundance and anti-citrullinated peptide antibody (ACPA) level determined using the robust LOWESS (locally weighted scatterplot smoothing) method. The shaded area indicates the range of the 94% confidence interval (from 3rd to 97th percentiles) through bootstrap resamplings.

**Figure 3** Metabolic shifts relate to the gut microbiome. (A-B) PLS-DA plots of fecal and serum metabolome. (C) Enrichment analysis reveals perturbation of metabolic pathways in feces and serum response to RA. Trp, tryptophan; Phe, phenylalanine; Tyr, tyrosine. (D) Total bile acid

content in fecal and serum samples. (E-F) Components of bile acids in feces (E) and serum (F). (G) Proportion of conjugated primary bile acids and conjugated secondary bile acids. (H) Serum concentration of FGF-19. (I) Average abundance of genera expressing specific enzymes involved in bile acid synthesis. (J) Per cents of bile acids in feces and serum. (K) Fecal and serum concentration of LCA. * $p< 0.05$, ** $p <0.01$, $^{ns}$not significant. BA, bile acid; 12-oxoLCA, 12-ketolithocholic acid; 3-DCA, 3-deoxycholic acid; 3-oxoCA, 3-dehydrocholic acid; 3-oxoCDCA, dehydrochenodeoxycholic acid; 3-oxoDCA, 3-oxodeoxycholic acid; 3-oxoLCA, dehydrolithocholic acid; 6-oxoLCA, 6-ketolithocholic acid; 7,12-dioxoLCA, 7,12-diketolithocholic acid; 7-oxoCA, 7-ketodeoxycholic acid; 7-oxoLCA, 7-ketolithocholic acid; ACA, allocholic acid; CA, cholic acid; CA-7S, cholic acid 7-sulfate; CDCA, chenodeoxycholic acid; CDCA-24Gln, chenodeoxycholic acid 24-acyl-β-d-glucuronide; DCA, deoxycholic acid; DHCA, dehydrocholic acid; GCA, glycocholic acid; GCDCA, glycochenodeoxycholic acid; GDCA, glycodeoxycholic acid; GHCA, glycohyocholic acid; GLCA, glycolithocholic acid; GUDCA, glycoursodeoxycholic acid; HCA, hyocholic acid; HDCA, hyodeoxycholic acid; IALCA, alloisolithocholic acid; isoCA, isocholic acid; isoDCA, isodeoxycholic acid; isoLCA, isolithocholic acid; isoUDCA, isoursodeoxycholic acid; LCA, lithocholic acid; LCA-3S, lithocholic acid-3-sulfate; muroCA, murocholic acid; NorCA, norcholic acid; NorDCA, 23-nordeoxycholic acid; TCA, taurocholic acid; TCA-3S, taurocholic acid 3-sulfate; TCDCA, taurochenodeoxycholic acid; TDCA, taurodeoxycholic acid; THDCA, taurohyodeoxycholic acid; TLCA, taurolithocholic acid; TLCA-3S, taurolithocholic acid-3-sulfate; TUDCA, tauroursodeoxycholic acid; UDCA, ursodeoxycholic acid; αMCA, α-muricholic acid; βMCA, β-muricholic acid; ωMCA, ω-muricholic acid; BSH, bile salt hydrolase; HSDH, hydroxysteroid dehydrogenase.

**Figure 4** Gut microbiome and microbial bile acid shifts reflect antigenic response in RA patients. (A) Correlations between RA-associated genera and clinical indices calculated by boosted additive generalized linear models. (B) Loess curves showed the correlation between ACPA and the genera involved in bile acid metabolism. (B) Heatmap showed that the dysregulation of bile acid metabolism related to RA phenotypes. In every rectangular lattice, the up and down triangles denoted the association between clinical indices and bile acid concentrations in fecal and serum samples, respectively. The left and right triangles represented the association between clinical indices and the percentages of bile acid in fecal and serum samples, respectively. In (A) and (C), $^+p < 0.1$, $*p < 0.05$, $**p < 0.01$, $***p < 0.001$. Red and blue colors denoted positive and negative associations, respectively. Grey color represented that bile acids were not detected in samples or excluded according to the quality control. ACPA, anti-citrullinated peptide antibody; BSH, bile salt hydrolase; CRP, C-reactive protein; RF, rheumatoid factor; DAS, disease activity score; ESR, erythrocyte sedimentation rate; HSDH, hydroxysteroid dehydrogenase.

**Figure 5** Microbial bile acids contribute to elevated ACPA expression and bone erosion in RA. (A-B) Shannon diversity (A) and beta diversity (Jaccard distance, B) of the two groups at species level. (C-D) The microbiota composition at genus level (C) and species level (D). The data are expressed as the relative abundance. (E) Relative abundance of *Bacteroides* species and *Clostridium scindens*. (F) Microbial bile acid transformations. Red color indicates bile acids increased in ACPA+ RA group, whereas blue color represents bile acids decreased.

Comparisons of ACPA+ RA with UC (+), HC (*), ACPA- RA (#) were shown. (G-I) Serum concentration of FGF-19 (G), PAD4 (H) and PAD2 (I). V1, validation set 1; V2, validation set 2. (J) Radiological evaluation of the hands from ACPA- RA group (n = 8) and ACPA+ RA group (n = 38). *$p<0.05$, **$p<0.01$, ***$p<0.001$, $^{ns}$not significant. (K) The representative hand X-ray radiographs with a disease duration of 3 years.

**Figure 6** Mediation analysis identifies linkages between Bacteroides species, bile acids and RA phenotypes. (A) Parallel coordinates chart showing the mediation effects of bile acids in feces and serum that were significant at $p < 0.05$. Shown are microbial species (left), bile acids (middle) and RA phenotypes (right). The curved lines connecting the panels indicate the mediation effects, with colors corresponding to different bile acids. (B) Analysis of the effect of *B. fragilis* on the level of ACPA and PAD2 mediated by bile acids. (C) Analysis of the effect of *B. thetaiotaomicron* on the level of ACPA mediated by isoLCA. (D) Analysis of the effect of *C. scindens* on the level of ACPA and SHS mediated by bile acids. In (B-D), orange and blue nodes represented factors excess and less in ACPA+ RA group respectively.

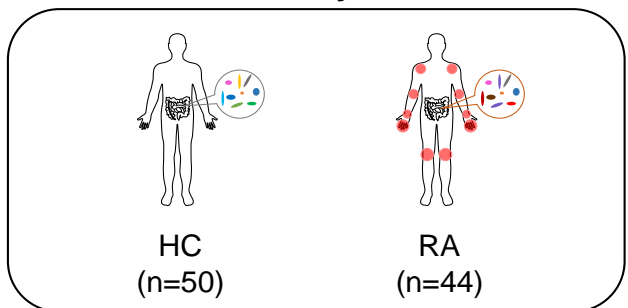
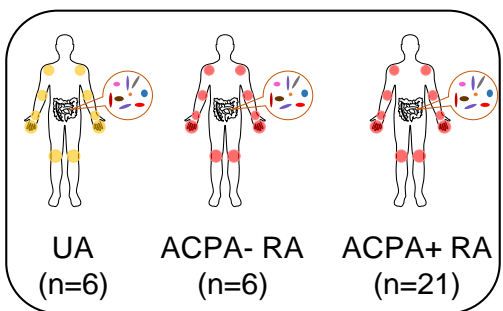
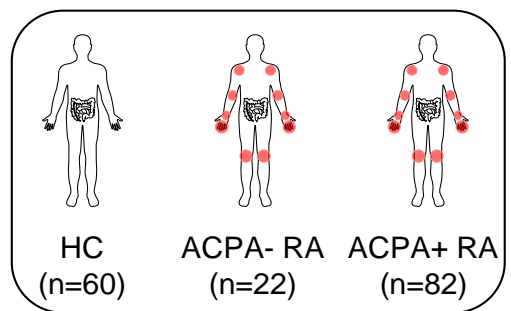

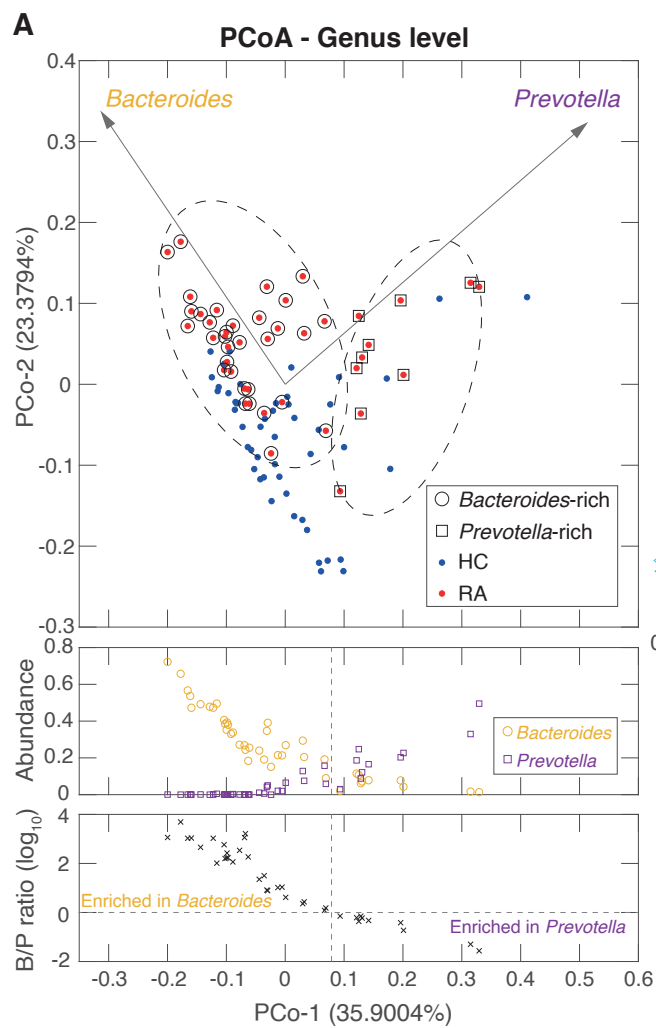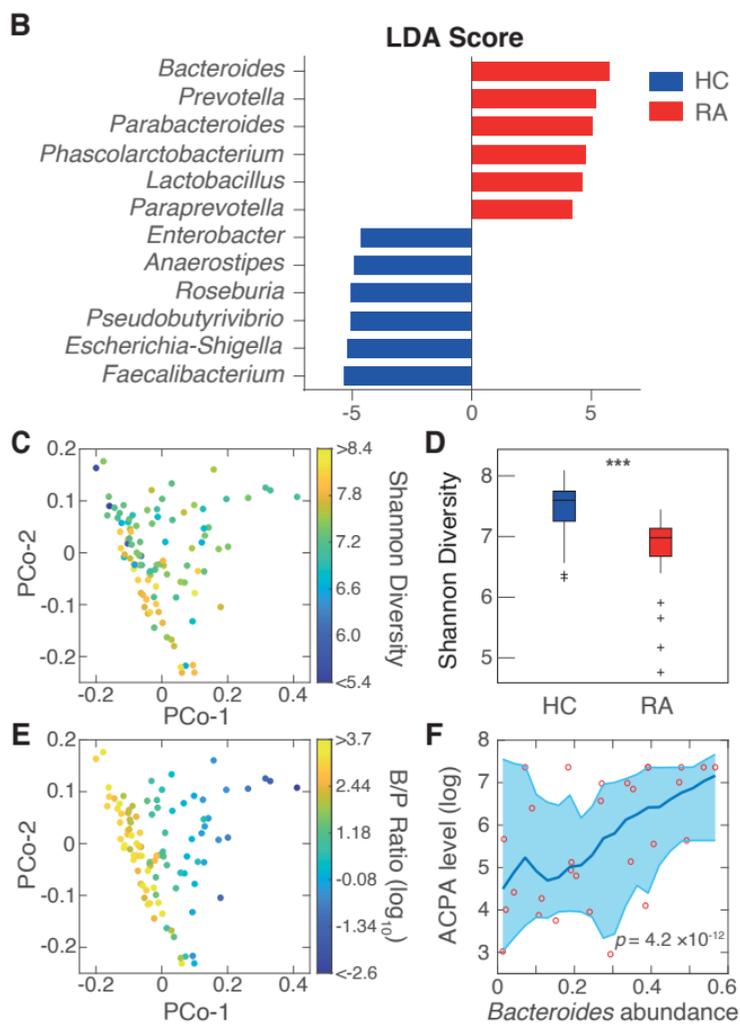

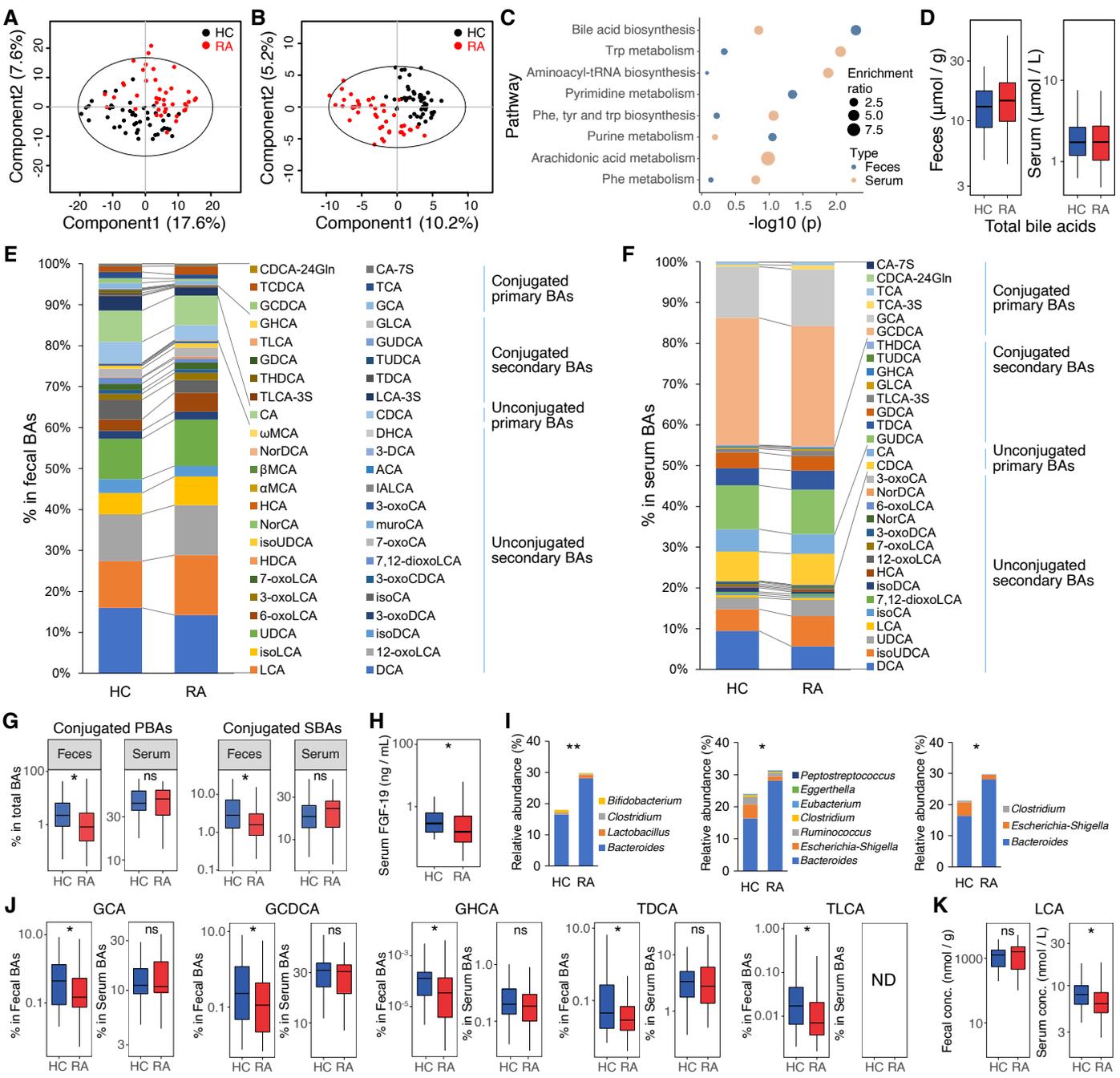

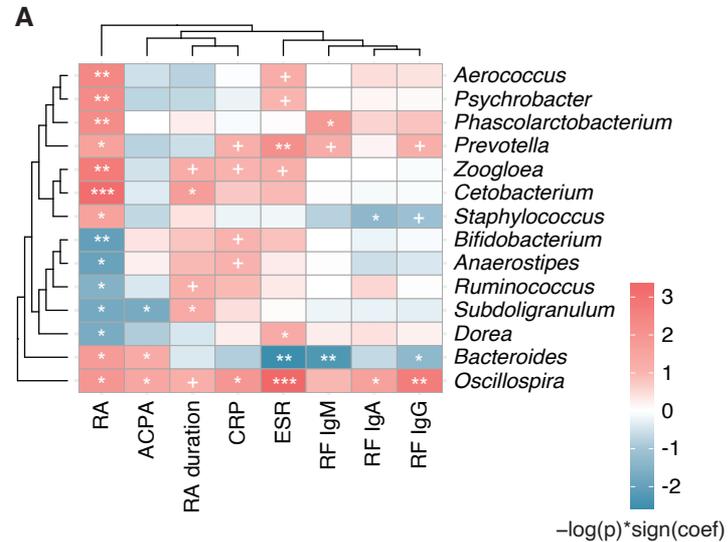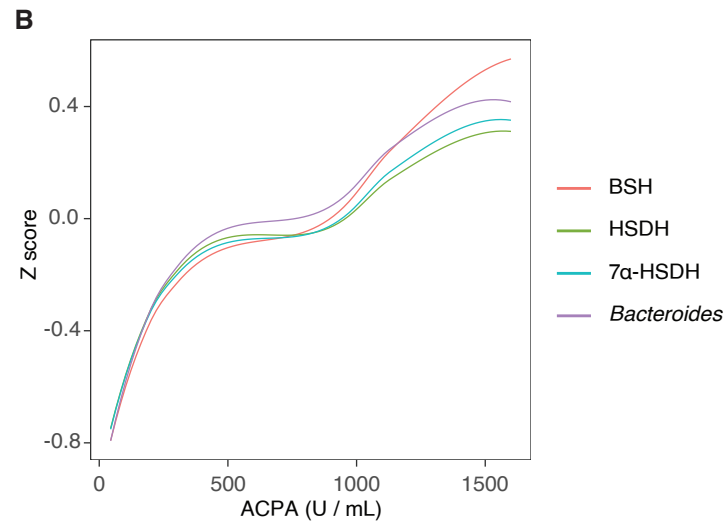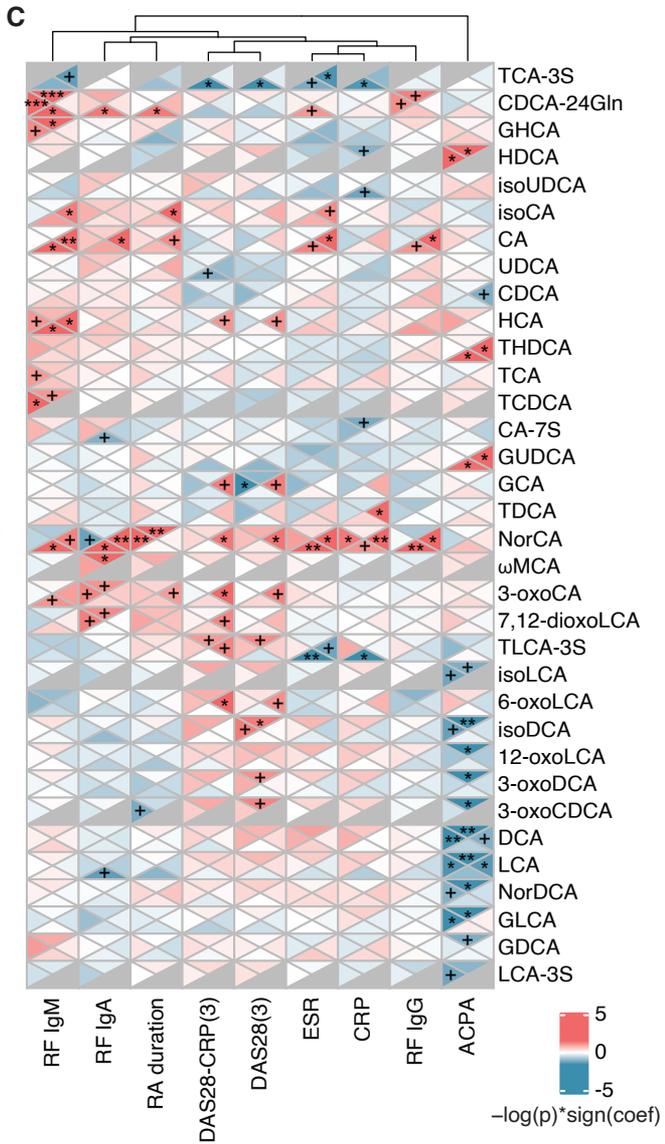

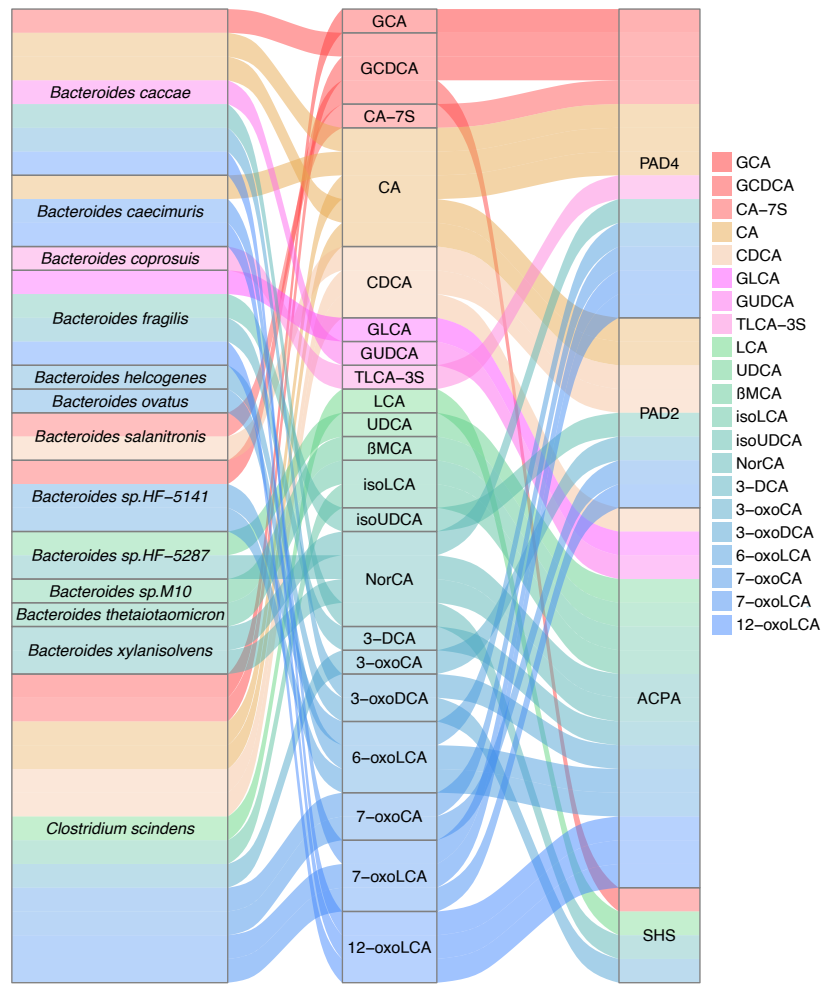
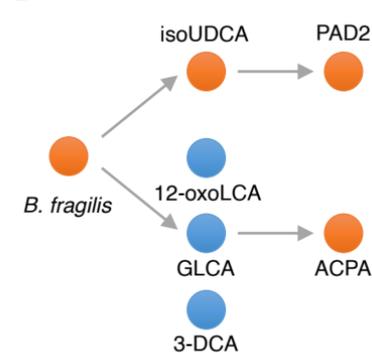
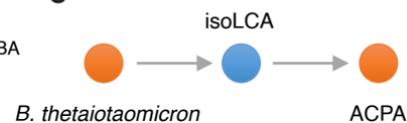
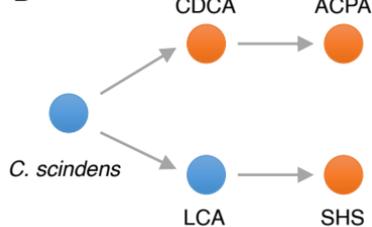